\documentclass[twocolumn,amsmath,amssymb,epsfig,aps]{revtex4}

\usepackage{graphicx}
\usepackage{dcolumn}
\usepackage{bm}
\usepackage{epsfig}

\def\fun#1#2{\lower3.6pt\vbox{\baselineskip0pt\lineskip.9pt
  \ialign{$\mathsurround=0pt#1\hfil##\hfil$\crcr#2\crcr\sim\crcr}}}

\input epsf

\def\be{\begin{equation}}
\def\ee{\end{equation}}
\def\ba{\begin{eqnarray}}
\def\ea{\end{eqnarray}}

\begin{document}

\title{Velocities as a probe of dark sector interactions}

\author{Kazuya Koyama, Roy Maartens, Yong-Seon Song }

\affiliation{Institute of Cosmology \& Gravitation, University of
Portsmouth, Portsmouth PO1~3FX, UK }

\date{\today}

\begin{abstract}

Dark energy in General Relativity is typically non-interacting
with other matter. However, it is possible that the dark energy
interacts with the dark matter, and in this case, the dark matter
can violate the universality of free fall (the weak equivalence
principle). We show that some forms of the dark sector interaction
do not violate weak equivalence. For those interactions that do
violate weak equivalence, there are no available laboratory
experiments to probe this violation for dark matter. But cosmology
provides a test for violations of the equivalence principle
between dark matter and baryons -- via a test for consistency of
the observed galaxy velocities with the Euler equation.

\end{abstract}

\maketitle

\section{Introduction}

Dark matter is currently only detected via its gravitational
effects, and there is an unavoidable degeneracy between dark
matter and dark energy within General Relativity. There could be a
hidden non-gravitational coupling between dark matter and dark
energy, and thus it is interesting to develop ways of testing for
such an interaction (see \cite{Friedman:1991dj, Gradwohl:1992ue, Bean:2008ac}
for earlier attempts).

One signal of a dark sector interaction could be a violation of
the weak equivalence principle (universality of free fall) by dark
matter, under the non-gravitational drag due to coupled dark
energy. Since Galileo shattered the myth that heavier objects fall
faster, the universality of free fall has been established as a
fundamental principle of gravity. Laboratory tests have been made
to show the independence of the acceleration of objects from their
masses and chemical composition. However, these tests apply to
baryonic matter, and no direct probe of dark matter is available.

If the interacting dark sector couples non-gravitationally to
baryonic matter, then existing laboratory tests provide
constraints on the dark sector interaction~\cite{Bovy:2008gh}.
Here we assume that there is zero (or negligible)
non-gravitational coupling between the dark sector and
standard-model fields. A difference in the acceleration between
dark matter and baryons could show up in the stellar distribution
in tidal trails of satellite galaxies~\cite{Kesden:2006vz}. This
same difference should also show up as an inconsistency when
interpreting the relation between galaxy peculiar velocities and
overdensities, as we explain below.

We assume that gravity on all scales is described by General
Relativity. Thus there is no gravitational mechanism to violate
the weak equivalence principle. Note that this is also true of
scalar-tensor theories, since the gravitational scalar degree of
freedom couples equally to all types of matter. Indeed, most
metric theories of modified gravity also respect the weak
equivalence principle (see, e.g.,~\cite{Sotiriou:2007zu}). Various
tests have been developed to discriminate between metric theories
of modified gravity, and non-interacting dark energy models in
General Relativity (see, e.g.,~\cite{mg}). But these tests do not
in general apply to dark energy that interacts with dark matter,
since a dark sector interaction can introduce new
degeneracies~\cite{Wei:2008vw}. We confine ourselves to the
question of how galaxy peculiar velocities can be used to detect
dark sector interactions within General Relativity.

\section{Interacting Dark Energy}

We briefly review the necessary background on perturbations of
interacting dark energy models in General Relativity. (For recent
work with further references, see, e.g.,~\cite{Valiviita:2008iv}.)

A general dark sector coupling may be described in the background
by the energy balance equations of cold dark matter ($c$) and dark
energy ($x$),
\begin{eqnarray}
  \label{relation}
\rho_c'  &=& - 3{\cal H}\rho_c+aQ_c\,,\label{cc}\\
\rho_{x}' &=& - 3{\cal H}(1+w_x)\rho_{x}+ aQ_x\,, ~~ Q_x=-Q_c\,,
\label{kg1}
\end{eqnarray}
where $w_x=P_x/\rho_x$, ${\cal H}=d\ln a/d\tau$ and $\tau$ is
conformal time, with $ds^2=a^2(-d\tau^2+d\vec{x}^{\,2}\,)$.  Here
$Q_c,Q_x$ are the rates of energy density transfer to dark matter
and energy respectively. In order to avoid stringent
``fifth-force" constraints, we assume that baryons ($b$), photons
($\gamma$) and neutrinos ($\nu$) are not coupled to dark energy
and are separately conserved.

In the Newtonian gauge the perturbed metric is given by
\begin{equation}
 ds^2=a^2 \Big[-(1+2\Psi)d\tau^2+(1-2\Psi)d\vec{x}\,^2 \Big] \,,
\end{equation}
where we have neglected anisotropic stress since we are interested
in the late universe. The total (energy-frame) four-velocity is
 \be \label{totu}
u^\mu = a^{-1}\Big(1-\Psi, \partial^i v \Big),
 \ee
where the velocity potential $v$ is defined by
 \be\label{vef}
(\rho+P) v=\sum (\rho_A+P_A) v_A\,,
 \ee
and $A=c,x,b,\gamma,\nu$. The $A$-fluid four-velocity is
 \be\label{ua}
u^\mu_A = a^{-1}\Big(1-\Psi, \partial^i v_A \Big).
 \ee
The covariant form of energy-momentum transfer is
\begin{equation}
\label{eqn:energyexchange} \nabla_\nu T^{\mu\nu}_{A } = Q^\mu_{A
}\,,
\end{equation}
where $Q^\mu_A=0$ for $A=b,\gamma,\nu$ in the late universe, while
$Q_c^\mu=-Q_x^\mu \neq 0$. The energy-momentum transfer
four-vector can be split relative to the total four-velocity as
 \be
Q_A^\mu =Q_A u^\mu + F_A^\mu\,,~~ Q_A=\bar{Q}_A + \delta Q_A\,, ~~
u_\mu F_A^\mu=0\,,
 \ee
where $Q_A$ is the energy density transfer rate and $F_A^\mu$ is
the momentum density transfer rate, relative to $u^\mu$. Then it
follows that $F_A^\mu=a^{-1}(0,\partial^if_A)$, where $f_A$ is a
momentum transfer potential, and
\begin{eqnarray}
Q^{A }_0 &=& -a\left[ Q_A(1+\Psi) + \delta Q_A
\right],\label{Qenergy}
\\Q^{A }_i
& = & a\partial_i\left( f_A+ Q_A v \right). \label{Qmomentum}
\end{eqnarray}
In the background, the energy-momentum transfer four-vectors have
the form $Q^\mu_{c}= a^{-1}(Q_c,\vec 0\,) =-Q^\mu_x\,,$ so that
there is no momentum transfer.

The evolution equations for the dimensionless density perturbation
$\delta_A=\delta\rho_A/\rho_A$ and for the velocity perturbation
are:
 \ba
&&\delta_A'+3{\cal H}(c_{sA}^2-w_A)\delta_A -(1+w_A)k^2v_A \nonumber\\
&&~~- 3{\cal H}\big[3{\cal H}(1+w_A)(c_{sA}^2-w_A)+w_A' \big] v_A
\nonumber\\
&&~~-3(1+w_A)\Psi' ={a\over \rho_A}\, \delta Q_A \nonumber\\
&&~~+{aQ_A \over \rho_A}\left[ \Psi-\delta_A-3{\cal
H}(c_{sA}^2-w_A) v_A \right]
\,,\label{dpa}\\
&& v_A'+{\cal H}\big(1-3c_{sA}^2\big)v_A+ {c_{sA}^2 \over
(1+w_A)}\,\delta_A +\Psi
\nonumber \\
&&~= {a\over (1+w_A)\rho_A}\Big\{Q_A \big[ v- (1+c_{sA}^2)v_A
\big]+ f_A\Big\}\!, \label{vpa}
 \ea
where $w_c=0=c_{sc}^2$ and $c_{sx}^2=1$.

For our purposes, we are interested in the behaviour of dark
matter in the Newtonian regime on sub-Hubble scales. In this case,
the perturbed continuity and Euler equations reduce to
 \ba
\delta_c'-k^2v_c &=& {a\over \rho_c}\left( \delta Q_c-Q_c
\delta_c\right), \label{dpc} \\ v_c'+{\cal H}v_c+ \Psi &=& {a\over
\rho_c}\left[Q_c(v-v_c)+f_c \right],\label{vpc}
 \ea
If the right-hand side of the continuity equation~(\ref{dpc}) is
nonzero, then the interaction will lead to a bias in the linear
regime between dark matter and baryons~\cite{Amendola:2001rc},
since the baryon overdensities obey
 \be
\delta_b'-k^2v_b=0\,.
 \ee
If the right-hand side of the Euler equation~(\ref{vpc}) is
nonzero, then the dark matter no longer follows geodesics and
breaks the weak equivalence principle, unlike baryons, for which
 \be
v_b' +{\cal H} v_b+ \Psi=0 \,.
 \ee

In the Newtonian regime, the Poisson equation becomes
 \be \label{poiss}
k^2\Psi=-4\pi G a^2\left(\rho_c \delta_c + \rho_b \delta_b
\right)\,.
 \ee
Here we neglect dark energy clustering, assuming that the sound
velocity of dark energy perturbations is $c_{sx}=1$. Dark energy
perturbations can be important on large scales depending on the
strength of interactions but they
are not important on sub-horizon scales as long as the sound velocity
of dark energy perturbations is positive -- since in that case, the gradient term
in the evolution equation for $\delta_x$ [see Eq.~(\ref{vpa})]
always dominates over the interaction terms. The evolution
equation for $\delta_c$ is then given by
\ba \label{deltaceq}
&& \delta_c''+{\cal H} \delta_c' - 4 \pi G a^2 \rho_c \delta_c
-{\cal H} \frac{a}{\rho_c}(\delta Q_c -Q_c \delta_c) \nonumber\\
&&{} -\Big[\frac{a}{\rho_c}(\delta Q_c -Q_c \delta_c)\Big]'
 -\frac{a}{\rho_c}k^2 \Big[
Q_c (v-v_c) +f_c \Big]=0. \nonumber\\
\ea

\section{Different types of interaction}

There is no fundamental theory that determines the form of the
interaction, i.e., of $Q_c^\mu$, so we are forced to use
phenomenological models. Here we consider three types of
interaction, each illustrated with a particular form: interactions
that do not change the continuity or Euler equations; interactions
that change only the Euler equation; interactions that change only
the continuity equation. The general case, where both equations
are modified, can be thought of as a linear superposition of the
last two cases.

\subsection{Continuity and Euler equations unchanged}

A general class of interactions may be defined by requiring that
there is no momentum exchange in dark matter rest frame,
 \be
Q_c^\mu = Q_c u_c^\mu\,,\label{gen}
 \ee
where $Q_c$ remains to be specified. For this class, we find from
Eqs.~(\ref{Qenergy}) and (\ref{Qmomentum}) that, for any $Q_c$, we
have $f_c =Q_c(v_c-v)$. Thus Eq.~(\ref{vpc}) becomes
 \be \label{vc1}
v_c'+{\cal H}v_c+ \Psi =0\,.
 \ee
This is the same Euler equation as the non-interacting case, so
that the dark matter velocity is not directly affected by the
interaction and there is no violation of weak equivalence. The
dark matter continues to follow geodesics, and feels no direct
drag force from the dark energy.

An example in the form of Eq.~(\ref{gen})
is~\cite{Valiviita:2008iv,Boehmer:2008av, Majerotto:2009np, Valiviita:2009nu}
 \be\label{C}
Q^\mu_{c}=  -\Gamma \rho_c\, u_c^\mu \,,
 \ee
where $\Gamma$ is a constant interaction rate. In this case
$Q_c=-\Gamma\rho_c(1+\delta_c)$ and Eq.~(\ref{dpc}) becomes
 \be
\delta_c'-k^2v_c=0\,.
\label{con0}
 \ee
The continuity equation is therefore the same as in the
non-interacting case.

Thus for this form of interaction, there is no violation of the
weak equivalence principle by dark matter, and no bias is induced
by the interaction. In fact, in the Newtonian regime, the only
signal of the dark sector interaction in structure formation to
linear order is via the modification of the background expansion
history. The evolution equation~(\ref{deltaceq}) for $\delta_c$
becomes
 \be \label{deltacun}
\delta_c''+{\cal H} \delta_c' - 4 \pi G a^2 (\rho_c \delta_c +
\rho_b \delta_b)=0,
 \ee
which is the same as in the uncoupled case. Thus the only imprint
of the dark sector interaction on $\delta_c$ is via the different
background evolution of ${\cal H}$ and $\rho_c$.

\subsection{Continuity equation modified}

If we keep Eq.~(\ref{gen}) but generalize Eq.~(\ref{C})
to~\cite{CalderaCabral:2008bx,CalderaCabral:2009ja}
 \be\label{Cgen}
Q^\mu_{c}= -(\Gamma_c \rho_c+ \Gamma_x \rho_x)\, u_c^\mu \,,
 \ee
then $\delta Q_c-Q_c\delta_c= \Gamma_x\rho_x(\delta_c-\delta_x)$.
Since dark energy does not cluster on sub-Hubble scales, we can
neglect the $\delta_x$ term, and we have
 \be \label{e2}
\delta_c'-k^2 v_c=a\Gamma_x {\rho_x \over \rho_c} \,\delta_c\,.
 \ee
For this interaction, the dark matter continues to follow
geodesics by virtue of Eq.~(\ref{vc1}), but the continuity
equation~(\ref{e2} is modified. As a consequence, there will be a
bias induced by the interaction.

The evolution equation~(\ref{deltaceq}) for $\delta_c$ becomes
 \ba
&& \delta_c''+\left({\cal H}
-a\Gamma_x \frac{\rho_x}{\rho_c}  \right)\delta_c'= 4\pi Ga^2\rho_b \delta_b \nonumber\\
&&~~{}+\Big [4\pi G a^2 \rho_c +2 a {\cal H}
\Gamma_x\frac{\rho_x}{\rho_c} +a \Gamma_x\Big(
\frac{\rho_x}{\rho_c}\Big)'
   \Big] \delta_c \,.\label{dc2gx}
 \ea
(This generalizes~\cite{CalderaCabral:2009ja}, where only the case
$\Gamma_c = 0$ is considered.)

The modification of the standard evolution for $\delta_c$ occurs
in 3 ways: firstly via the modified expansion history in the
background ${\cal H}$ and $\rho_c$; secondly by the modified
Hubble friction term ${\cal H} \to {\cal H}[1-a\Gamma_x
\rho_x/{\cal H}\rho_c ]$; and thirdly by the modified effective
gravitational coupling for dark matter -- dark matter particle
interactions,
 \be
G_{\rm eff}=G\Big[1+{{\cal H}\rho_x \over 2\pi G a \rho_c^2}+
{\Gamma_x \over 4\pi G a \rho_c}\Big({\rho_x \over \rho_c} \Big)'
\Big].
 \ee

\subsection{Euler equation modified}

A second general class of interactions has no momentum exchange in
the dark energy frame,
 \be
Q_c^\mu = Q_c u_x^\mu\,.\label{gen2}
 \ee
It follows that $f_c =Q_c(v_x-v)$, and hence
 \be \label{vcux}
v_c'+{\cal H}v_c+ \Psi = {a\over \rho_c}Q_c(v_x-v_c)\,.
 \ee
In this case, there is an explicit deviation of the dark matter
velocity relative to the non-interacting case. The dark matter no
longer follows geodesics in general. Note that, even though dark
energy does not cluster on sub-Hubble scales, we cannot in general
neglect the dark energy velocity $v_x$ relative to the dark matter
velocity $v_c$ in Eq.~\eqref{vcux}.

An example of the form of Eq.~\eqref{gen2}
is~\cite{Wetterich:1994bg}
 \be\label{stc}
Q^\mu_{c}=-\alpha \rho_c \nabla^\mu \varphi\,,
 \ee
where $\varphi$ is the scalar field that describes dark energy and
$\alpha$ is a coupling constant. Note that $\nabla^\mu \varphi$ is
parallel to the dark energy four-velocity $u_x^\mu$:
 \be \label{ux}
u_x^\mu={1\over a}\Big(1-\Psi, -{\partial^i\delta\varphi \over
\varphi'}\Big)\,, ~~ v_x=-{\delta\varphi \over \varphi'}\,.
 \ee
In this case, 
$Q_c = a^{-1}\alpha (\rho_c  \varphi' + \delta \rho_c \varphi' + \rho_c \delta\varphi' -\rho_c \varphi' \Psi )$. The perturbed
Klein-Gordon equation is~\cite{Hwang:2001fb}
 \ba
&& \delta\varphi'' +2{\cal H}\delta
\varphi'+(k^2+a^2V_{\varphi\varphi}) \delta\varphi \nonumber\\
&&~~{}= 2\varphi'(\Psi'+{\cal H} \Psi)+2\varphi''\Psi-\alpha
a^2\rho_c\delta_c\,,
 \ea
where $V(\varphi)$ is the quintessence potential. In the Newtonian
regime, the last term on the right dominates over the other terms,
while the $k^2$ term dominates on the left, leading to
 \be \label{dvphi}
k^2\delta\varphi = -\alpha a^2\rho_c\delta_c\,.
 \ee

It follows from Eqs.~\eqref{vcux}, (\ref{stc}) and \eqref{ux} that
 \be \label{vpc2}
v_c'+{\cal H}v_c+\Psi=-\alpha \varphi'\Big(v_c+{\delta\varphi
\over \varphi'}\Big),
 \ee
confirming the violation of weak equivalence. For the perturbed
continuity equation~\eqref{dpc}, the right-hand side becomes
$-\alpha\delta\varphi'$. By Eq.~\eqref{dvphi}, this term is
suppressed by $k^{-2}$ relative to the $\delta_c'$ term on the
left-hand side, and therefore to a good approximation we have
 \be
 \label{vc}
\delta_c'-k^2v_c=0\,.
 \ee
Using (\ref{dvphi}) and (\ref{vc}), the evolution
equation~(\ref{deltaceq}) for $\delta_c$ becomes
 \ba
&& \delta_c''+({\cal H}+\alpha\varphi')\delta_c'= 4\pi G a^2 \rho_b \delta_b \nonumber \\
&&~~{}+ 4\pi G a^2\Big(1+{\alpha^2 \over 4\pi G} \Big)
\rho_c\delta_c \,.  \label{dpc2}
 \ea

As in the case of Eq.~(\ref{dc2gx}), the modification of the
standard evolution for $\delta_c$ occurs in 3
ways~\cite{Amendola:2001rc}: firstly via the modified expansion
history in the background ${\cal H}$ and $\rho_c$; secondly by the
modified Hubble friction term ${\cal H} \to {\cal
H}[1+\alpha\varphi'/{\cal H}]$; and thirdly by the modified
effective gravitational coupling for dark matter -- dark matter
particle interactions,
 \be
G_{\rm eff}=G\Big(1+{\alpha^2 \over 4\pi G} \Big).
 \ee
These effects are incorporated in the modified $N$-body simulations for
this form of interacting dark energy~\cite{Maccio:2003yk}.

\section{Testing for dark sector interactions}

In this section, we discuss several possible ways to use
observations to constrain the dark sector interactions discussed
in the previous section.

\subsection{Continuity and Euler equations unchanged}

We first consider the case where there is no modification to the
dynamics of perturbations in the Newtonian regime. The difference
comes purely from the modified background evolution. If dark
matter interacts with dark energy, the dark matter density no
longer decays like $a^{-3}$. This affects the distance measures in
the Universe and thus changes the measurements of CMB, SNe and
Baryon Acoustic Oscillations. By combining these observations, we
can measure today's matter density and then determine the matter
energy density at the last scattering surface. However, the
distance is determined by integrating over the expansion history
and we cannot directly check the deviation at each redshift from
the standard behaviour, $\rho_c \propto a^{-3}$.

There is an independent way to measure the dark matter density
using structure formation. From the Poisson equation, the dark
matter density can be written as
\begin{equation} \label{om}
\omega_m(a) \equiv \Omega_m(a) h^2  = -\frac{2 \Psi(k, a)}{3
\delta_c (k, a)} \left( \frac{k h}{a H_0} \right)^2,
\end{equation}
where we neglected the baryon contribution for simplicity (we are
only illustrating the principle, rather than making quantitative
predictions). One way to measure $\delta_c$ is to reconstruct
$\delta_c$ from peculiar velocities using the continuity equation
Eq.~(\ref{con0}) because in this case there is no modification
to the continuity equation and no violation of weak equivalence
principle. On the other
hand, weak lensing measures directly $\Psi$ without bias. Thus we
can use Eq.~(\ref{om}) to predict the background evolution of
matter density from structure formation.

\begin{figure}[t]
\centerline{
\includegraphics[width=7cm]{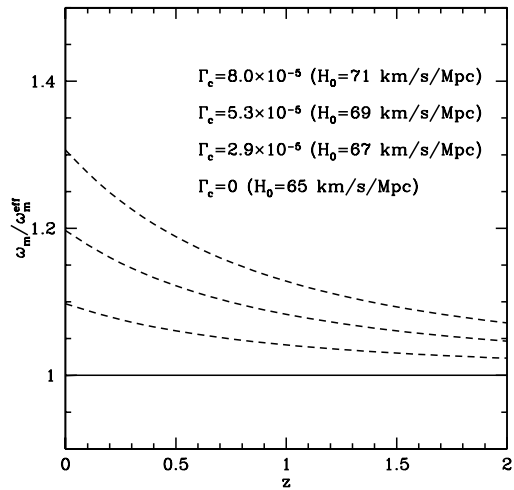}}
\caption{The ratio between the true matter density obtained from
structure formation and the density estimated from geometrical
tests assuming the non-interacting adiabatic behaviour $\rho_m
\propto a^{-3}$.} \label{fig:1}
\end{figure}

In Fig.~\ref{fig:1}, we plot $\omega_m/\omega_m^{\rm eff}$, where
$\omega_m$ is the true matter density measured by weak lensing,
and $\omega_m^{\rm eff}$ is derived from the background
measurement of $\omega_m$ at the last scattering surface, assuming
$\rho_m \propto a^{-3}$. At late times when the interaction
becomes important, this ratio deviates from 1 due to the
non-adiabatic decay of the dark matter density. In this way, we
can check the modification to the behaviour of the matter density
at each redshift, using tomographic measurements of $\Psi$ from
weak lensing.

\subsection{Test of the continuity equation}

In the case where the interaction changes only the continuity
equation, there is no difference between the peculiar velocities
of baryons and dark matter. We assume that galaxies can be treated
as test particles that are made of baryons and whose peculiar
velocities, $v_g$, are determined by baryon peculiar velocities.
Although there is an indication that this assumption is
valid~\cite{Percival:2008sh}, this should be tested by N-body
simulations carefully in the presence of interaction. We leave
this for a future work.

With this assumption, we can determine peculiar velocities of
baryons, $v_b$, from peculiar velocities of galaxies, $v_g$. The
latter can be measured for example by redshift-space distortions
(see~\cite{Song:2008qt,White:2008jy} for recent work). Then it is
possible to determine dark matter peculiar velocities because
$v_c= v_b$.

On the other hand, density perturbations can be measured from the
galaxy distribution with a knowledge of bias. One possibility to
measure bias is to use weak lensing. Weak lensing measures $\Psi$
without bias and $\delta_c$ can be derived from the Poisson
equation~(\ref{poiss}). Note that in order to measure $\Psi$ from
$\delta_c$, it is necessary to measure the true evolution of $\rho_c$,
which is modified by interactions. However, we found that the
modification to the continuity equation has significant effects even
in weak interactions cases where the effect of interactions on $\rho_c$ is
negligible. Thus in the following we only consider the case where we can
neglect the effect of interactions on the evolution of $\rho_c$. Another
possibility is to use the peculiar
velocity measurements. In the case that we consider here, the
Euler equation is not modified [see Eq.~(\ref{vc1})] and it is
possible to reconstruct $\Psi$ from $v_c$. Then again using the
Poisson equation, $\delta_c$ can be
derived~\cite{Hu:2003pt,Acquaviva:2008qp,Song09}. In this way we
can test whether the continuity equation is modified.

\begin{figure}[t]
\centerline{
\includegraphics[width=7cm]{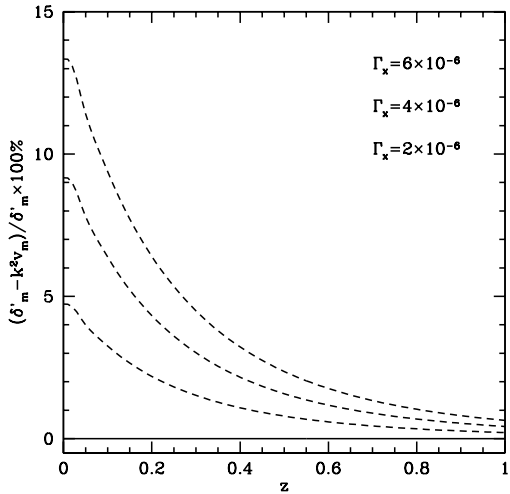}}
\caption{The breakdown of the continuity equation by an
interacting dark energy model. In this model, $v_b=v_c$ and
$\delta_c'-k^2 v_c = a \Gamma_x (\rho_x/\rho_c)\delta_c$.}
\label{fig:2}
\end{figure}

Fig.~\ref{fig:2} demonstrates the breakdown of the standard
continuity equation by an interacting dark energy model. We used a
model where $\Gamma_x \neq 0$ and $\Gamma_c=0$.

\subsection{Test of weak equivalence principle}

The weak equivalence principle is broken when the Euler equation
for dark matter is modified. In this case, there is a difference
between the peculiar velocities of dark matter and baryons. With
the assumption that galaxies trace baryon peculiar velocities, we
measure baryon peculiar velocities say from red-shift distortions.
Unlike the previous case, dark matter peculiar velocities are
different. However, without knowing that there is an interaction
between dark matter and dark energy, we {\it estimate} dark matter
peculiar velocity as
\begin{equation}
v_c^{\rm est} = v_b =v_g\,.
\end{equation}
The estimated peculiar velocity is different from true peculiar
velocity of dark matter: $v_c \neq v_c^{\rm est}$.

\begin{figure}[h]
\centerline{
\includegraphics[width=7cm]{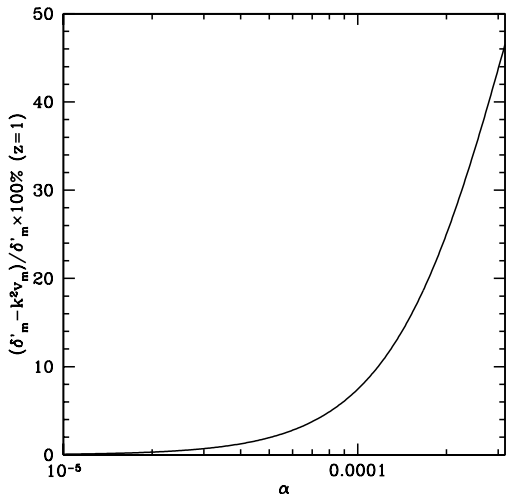}}
\caption{The breakdown of the weak equivalence principle for dark
matter. In this model, the continuity equation is not broken but
$v_c \neq v_b$.} \label{fig:3}
\end{figure}

If the continuity equation is not modified, as happens for the
model of Eq.~(\ref{stc}), then  the true peculiar velocity
satisfies the same continuity equation as the uncoupled case.
Thus, if we use the estimated peculiar velocity, the continuity
equation is {\it apparently} broken
\begin{equation}
\delta_c' - k^2 v_c^{\rm est} \neq 0.
\end{equation}
In this case, the continuity equation is not broken but $v_c \neq
v_b$. Then we can apply the same analysis as in the previous
section. We can measure $\delta_c$ from weak lensing. Then it is
possible to prove the breakdown of the weak equivalence principle
through the apparent breakdown of the continuity equation.
Fig.~\ref{fig:3} demonstrates this idea.

\section{Conclusions}

An interaction between dark matter and dark energy could exist in
various ways which are not detectable by any direct probe. We
investigated how the Euler equation and the continuity equation
for dark matter could be modified by such an interaction, taking
care to provide a covariant analysis of momentum transfer.
Modification of the Euler equation indicates a deviation of the dark
matter motion from geodesic, under the drag force of dark energy --
and a consequent breaking of the weak equivalence principle for
dark matter. Using three different forms of interaction as
examples, we considered interacting models in which:\\ (A) neither
the Euler nor continuity equations are modified, so that the
effect of the interaction in the Newtonian regime is purely via
the different background evolution;\\ (B)~the Euler equation is
unchanged but the continuity equation is modified (and
consequently a new bias is introduced by the interaction); \\
(C)~the continuity equation is unchanged but the Euler equation is
modified, leading to violation of weak equivalence.

We discussed how in principle observations could be used to detect these different forms of interacting dark energy. In case (A), we used the fact that the continuity and Euler equations are unchanged to devise a test based on the non-adiabatic redshifting of the dark matter. This test uses independent measurements of
the Newtonian potential and the density perturbation via the Poisson equation, to compute the true matter density and show that it deviates from the non-interacting case.

In cases (B) and (C), the
effects of the violation of the continuity equation or Euler
equation are stronger than the non-standard redshifting of the background matter density. We showed how, given a
knowledge of bias from weak lensing, tests could be constructed for the breakdown of the continuity or the Euler equation.

A further issue raised by our investigations is how to distinguish interacting dark energy from modified gravity. This is left for future work.

\begin{acknowledgements}
The authors are supported by the UK's Science \& Technology
Facilities Council. KK is supported by the European Research
Council and Research Councils UK.
\end{acknowledgements}


\begin{thebibliography}{99}

\bibitem{Friedman:1991dj}
  J.~A.~Frieman and B.~A.~Gradwohl,
  Phys.\ Rev.\ Lett.\  {\bf 67}, 2926 (1991).


\bibitem{Gradwohl:1992ue}
  B.~A.~Gradwohl and J.~A.~Frieman,
  Astrophys.\ J.\  {\bf 398}, 407 (1992).

\bibitem{Bean:2008ac}
  R.~Bean, E.~E.~Flanagan, I.~Laszlo and M.~Trodden,
  Phys.\ Rev.\  D {\bf 78}, 123514 (2008)
  [arXiv:0808.1105 [astro-ph]].



\bibitem{Bovy:2008gh}
  J.~Bovy and G.~R.~Farrar,
  arXiv:0807.3060 [hep-ph];\\
  S.~M.~Carroll, S.~Mantry, M.~J.~Ramsey-Musolf and C.~W.~Stubbs,
  arXiv:0807.4363 [hep-ph].

\bibitem{Kesden:2006vz}
  M.~Kesden and M.~Kamionkowski,
  Phys.\ Rev.\  D {\bf 74}, 083007 (2006)
  [arXiv:astro-ph/0608095];\\
  J.~A.~Keselman, A.~Nusser and P.~J.~E.~Peebles,
  arXiv:0902.3452 [astro-ph.GA].

\bibitem{Sotiriou:2007zu}
  T.~P.~Sotiriou, V.~Faraoni and S.~Liberati,
  Int.\ J.\ Mod.\ Phys.\  D {\bf 17}, 399 (2008)
  [arXiv:0707.2748 [gr-qc]].

\bibitem{mg}
  A.~Lue, R.~Scoccimarro and G.~Starkman,
  Phys.\ Rev.\  D {\bf 69}, 044005 (2004)
  [arXiv:astro-ph/0307034];\\
  L.~Knox, Y.~S.~Song and J.~A.~Tyson,
  Phys.\ Rev.\  D {\bf 74} (2006) 023512;\\
  M.~Ishak, A.~Upadhye and D.~N.~Spergel,
  Phys.\ Rev.\  D {\bf 74} (2006) 043513
  [arXiv:astro-ph/0507184];\\
  E.~V.~Linder,
  Phys.\ Rev.\  D {\bf 72}, 043529 (2005)
  [arXiv:astro-ph/0507263];\\
  E.~Bertschinger,
  Astrophys.\ J.\  {\bf 648}, 797 (2006)
  [arXiv:astro-ph/0604485];\\
  S.~Wang, L.~Hui, M.~May and Z.~Haiman,
  Phys.\ Rev.\  D {\bf 76}, 063503 (2007)
  [arXiv:0705.0165 [astro-ph]];\\
  W.~Hu and I.~Sawicki,
  Phys.\ Rev.\  D {\bf 76}, 104043 (2007)
  [arXiv:0708.1190 [astro-ph]];\\
  M.~A.~Amin, R.~V.~Wagoner and R.~D.~Blandford,
  arXiv:0708.1793 [astro-ph];\\
  B.~Jain and P.~Zhang,
  Phys.\ Rev.\  D {\bf 78}, 063503 (2008)
  [arXiv:0709.2375 [astro-ph]];\\
  Y.~Wang,
  JCAP {\bf 0805}, 021 (2008)
  [arXiv:0710.3885 [astro-ph]];\\
  E.~Bertschinger and P.~Zukin,
  arXiv:0801.2431 [astro-ph];\\
  S.~F.~Daniel, R.~R.~Caldwell, A.~Cooray and A.~Melchiorri,
  arXiv:0802.1068 [astro-ph];\\
  L.~Guzzo {\it et al.},
  Nature {\bf 451}, 541 (2008)
  [arXiv:0802.1944 [astro-ph]];\\
  Y.~S.~Song and K.~Koyama,
  JCAP {\bf 0901}, 048 (2009)
  [arXiv:0802.3897 [astro-ph]];\\
  Y.~S.~Song and O.~Dore,
  JCAP {\bf 0903} (2009) 025;\\
  G.~B.~Zhao, L.~Pogosian, A.~Silvestri and J.~Zylberberg,
  Phys.\ Rev.\  D {\bf 79}, 083513 (2009)
  [arXiv:0809.3791 [astro-ph]];\\
  G.~B.~Zhao, L.~Pogosian, A.~Silvestri and J.~Zylberberg,
  arXiv:0905.1326 [astro-ph.CO];\\
  P.~Zhang, R.~Bean, M.~Liguori and S.~Dodelson,
  arXiv:0809.2836 [astro-ph].


\bibitem{Wei:2008vw}
  H.~Wei and S.~N.~Zhang,
  Phys.\ Rev.\  D {\bf 78}, 023011 (2008)
  [arXiv:0803.3292 [astro-ph]].

\bibitem{Majerotto:2009np}
  E.~Majerotto, J.~Valiviita and R.~Maartens,
  arXiv:0907.4981 [astro-ph.CO].


\bibitem{Valiviita:2009nu}
  J.~Valiviita, R.~Maartens and E.~Majerotto,
  arXiv:0907.4987 [astro-ph.CO].

\bibitem{Valiviita:2008iv}
  J.~Valiviita, E.~Majerotto and R.~Maartens,
  JCAP {\bf 0807}, 020 (2008)
  [arXiv:0804.0232 [astro-ph]].

\bibitem{Amendola:2001rc}
  L.~Amendola and D.~Tocchini-Valentini,
  Phys.\ Rev.\  D {\bf 66}, 043528 (2002)
  [arXiv:astro-ph/0111535].

\bibitem{Boehmer:2008av}
  C.~G.~Boehmer, G.~Caldera-Cabral, R.~Lazkoz and R.~Maartens,
  Phys.\ Rev.\  D {\bf 78}, 023505 (2008)
  [arXiv:0801.1565 [gr-qc]].

\bibitem{CalderaCabral:2008bx}
  G.~Caldera-Cabral, R.~Maartens and L.~A.~Urena-Lopez,
  Phys.\ Rev.\  D {\bf 79}, 063518 (2009)
  [arXiv:0812.1827 [gr-qc]].

\bibitem{CalderaCabral:2009ja}
  G.~Caldera-Cabral, R.~Maartens and B.~M.~Schaefer,
  arXiv:0905.0492 [astro-ph.CO].

\bibitem{Wetterich:1994bg}
  C.~Wetterich,
  Astron.\ Astrophys.\  {\bf 301}, 321 (1995)
  [arXiv:hep-th/9408025];\\
  L.~Amendola,
  Phys.\ Rev.\  D {\bf 60}, 043501 (1999)
  [arXiv:astro-ph/9904120];\\
  D.~J.~Holden and D.~Wands,
  Phys.\ Rev.\  D {\bf 61}, 043506 (2000)
  [arXiv:gr-qc/9908026].

\bibitem{Hwang:2001fb}
  J.~c.~Hwang and H.~Noh,
  Class.\ Quant.\ Grav.\  {\bf 19}, 527 (2002)
  [arXiv:astro-ph/0103244].

\bibitem{Maccio:2003yk}
  A.~V.~Maccio, C.~Quercellini, R.~Mainini, L.~Amendola and S.~A.~Bonometto,
  Phys.\ Rev.\  D {\bf 69}, 123516 (2004)
  [arXiv:astro-ph/0309671];\\
  M.~Baldi, V.~Pettorino, G.~Robbers and V.~Springel,
  arXiv:0812.3901 [astro-ph].

\bibitem{Percival:2008sh}
  W.~J.~Percival and M.~White,
  arXiv:0808.0003 [astro-ph].

\bibitem{Song:2008qt}
  Y.~S.~Song and W.~J.~Percival,
  arXiv:0807.0810 [astro-ph].

\bibitem{White:2008jy}
  M.~White, Y.~S.~Song and W.~J.~Percival,
  arXiv:0810.1518 [astro-ph].

\bibitem{Hu:2003pt}
  W.~Hu and B.~Jain,
  Phys.\ Rev.\  D {\bf 70} (2004) 043009
  [arXiv:astro-ph/0312395].

\bibitem{Acquaviva:2008qp}
  V.~Acquaviva, A.~Hajian, D.~N.~Spergel and S.~Das,
  Phys.\ Rev.\  D {\bf 78} (2008) 043514
  [arXiv:0803.2236 [astro-ph]].

\bibitem{Song09}
  Y-S.~Song, C.~Sabiu and R.~Nichol and C.~Miller,
  {\it submmited to JCAP} (2009).




\end{thebibliography}
\end{document}